\documentclass{article}
\usepackage{amsmath}
\usepackage{amssymb}
\usepackage{listings}
\usepackage{xcolor}
\usepackage{graphicx}
\usepackage{color}
\usepackage{xcolor}
\usepackage[all]{xy}
\usepackage{microtype}
\usepackage{multicol}
\usepackage{latexsym}
\usepackage{makeidx}
\usepackage{float}
\usepackage{wallpaper}
\usepackage{listings}
\usepackage{xcolor}
\usepackage[T1]{fontenc}
\usepackage{textcomp}
\usepackage[space=true]{accsupp}

\graphicspath{ {imagenes-bloque-1/} }
\definecolor{rojo}{rgb}{0.9451,0.58039,0.54118}
\definecolor{codegray}{rgb}{0.5,0.5,0.5}
\definecolor{naranja}{rgb}{ 0.12941, 0.18431, 0.23529}
\definecolor{backcolour}{rgb}{0.95686, 0.96471, 0.96471}
\definecolor{azure(colorwheel)}{rgb}{0.20392,0.59608,0.85882}
\definecolor{morado}{rgb}{0.4902,0.23529,0.59608}
\definecolor{verde}{rgb}{
0.15686,0.70588,0.38824}
\definecolor{Orange}{rgb}{0.90196,0.49412,0.13333}
\definecolor{cafe}{rgb}{0.79608,0.09804,0.09804}
\lstdefinestyle{mystyle}{language=python,                    backgroundcolor=\color{backcolour},   
    commentstyle=\color{rojo},
    keywordstyle=\color{azure(colorwheel)}, 
    numberstyle=\tiny\color{codegray},
    stringstyle=\color{verde},
    basicstyle=\color{naranja}\ttfamily\footnotesize,
    breakatwhitespace=false,       
    breaklines=true,               captionpos=b,                  keepspaces=true,               numbers=left,                  numbersep=5pt,                 showspaces=false,              
    showstringspaces=false,
    showtabs=true,              tabsize=2,
    literate=         {á}{{\'a}}1
         {é}{{\'e}}1
         {í}{{\'i}}1
         {ó}{{\'o}}1
         {ú}{{\'u}}1
         {0}{{{\color{Orange}0}}}1
    {1}{{{\color{Orange}1}}}1
    {2}{{{\color{Orange}2}}}1
    {3}{{{\color{Orange}3}}}1
    {4}{{{\color{Orange}4}}}1
    {5}{{{\color{Orange}5}}}1
    {6}{{{\color{Orange}6}}}1
    {7}{{{\color{Orange}7}}}1
    {8}{{{\color{Orange}8}}}1
    {9}{{{\color{Orange}9}}}1
    {*}{{\char42}}1
    {-}{{\char45}}1
    {\ }{{\copyablespace}}1,
    upquote=true,
    columns=fullflexible,
    emph=[1]{incmatrix, mueller_aproximation , matrix_norm , appr_mueller_matrix,van_der_mee_theorem,k_irreducible, k_primitive, request_matrix, ecm, make_invertible,  appr_invert_mueller_matrix, appr_k_primitive},
    emphstyle=[1]{\color{morado}\ttfamily},
     emph=[2]{as},
    emphstyle=[2]{\color{azure(colorwheel)}\ttfamily},
    emph=[3]{category, figsize, projection, decimals, cmap, linewidth, antialiased, shrink, aspect, location, precision, suppress, ha, fontsize, category},
    emphstyle=[3]{\color{cafe}\ttfamily},
     emph=[4]{True, False},
    emphstyle=[4]{\color{Orange}\ttfamily},
}
\newcommand{\copyablespace}{\BeginAccSupp{method=hex,unicode,ActualText=00A0}\hphantom{x}\EndAccSupp{}}
\definecolor{azul}{RGB}{66,212,244}

\input{tcilatex}

\begin{document}

\begin{center}
\textbf{THE CONE OF THE MUELLER MATRICES}

\textbf{J. Iv\'{a}n L\'{o}pez-Reyes, J. Oth\'{o}n Parra-Alcantar, }

\textbf{and} \textbf{Martha Y. Takane}$^{(\ast )}$

\textbf{Instituto de Matem\'{a}ticas }

\textbf{Universidad Nacional Aut\'{o}noma de M\'{e}xico (UNAM)}

\textbf{\noindent }
\end{center}

\noindent $(\ast )${\small \ Corresponding author: takane@im.unam.mx }

\noindent {\small https://www.matem.unam.mx/fsd/takane}

\bigskip

\noindent \textbf{Abstract.} In the study of polarized light, there are two
basic notions: the Stokes vectors and the matrices which preserve them,
called Mueller matrices. The set of Stokes vectors forms a cone: the Future
Light Cone. In this work we will see that the Mueller matrices also form a
cone in the vector space of real matrices of size $4\times 4$, called the
Mueller Cone. We obtain some properties of the Mueller cone, which in turn
will be translated into properties of the Stokes vectors. As an application
we will give a computational program to calibrate polarimeters by means of
the eigenvectors of Mueller matrices (ECM). We also include programs to see
if a matrix is {}{}Mueller, to approximate a matrix by a Mueller matrix, to
approximate it by an invertible Mueller matrix and by a
Stokes-cone-primitive Mueller matrix.

\textbf{\bigskip }


\bigskip

\noindent \textbf{Introduction}{\large .} A cone $K$ in a normed real vector
space is a closed set under nonnegative linear combinations, limits (when
they exist) and the only vector in $K$ whose additive inverse is also in $K$
is the zero vector. In the last century, Cone Theory has been a very useful
tool in Economics, see Nikaido [N68]. In recent years, this theory has been
applied to different sciences, mainly through the Birkhoff-Vandergraft
Theorem, see [Bi67] and [V68]. This theorem is a generalization in terms of
cones of the well-known Perron-Frobenius Theorem. For the benefit of the
reader, we have added an appendix with the basics of cone theory. The set of
the Stokes vectors is a cone that we will call the Stokes cone that has
already been used by Givens and Kostinski [GK93]. The Stokes cone is exactly
the Future Light Cone. Here we will prove that the set of (realizable)
Mueller matrices is also a cone and we will see the utility of seeing it as
a cone. For example, properties of the Mueller cone can give properties of
the Stokes vectors and give alternative proofs of known results about them.
As an application we will give an Eigenvalue Calibration Method of a
polarimeter, in the sense of Compain, Poirier and Drevillon, see [CPD99].

We include computational programs to 1. Deduce if a matrix is a Mueller
matrix, 2. Give an approximation of a matrix by a Mueller matrix, 3. An
approximation of a Mueller matrix by Mueller invertibles, 4. An
approximation of a Mueller matrix by a Stokes-cone-primitive Mueller matrix,
see (4.7) and 5. An Eigenvalue Calibration Method. All these programs and
implementations can be found in

\noindent \textbf{%
https://github.com/IvanLopezR22/cone-of-mueller-matrices/tree/master}

\bigskip

Throughout this article, $\mathbb{R}$ and $\mathbb{C}$ denote the fields of
the real and complex numbers with the topology given by their usual norms,
see in the Appendix the basic notions of Cone theory.

\bigskip

\section{\noindent \textbf{The} \textbf{Stokes cone}}

The Stokes cone is the set of the Stokes vectors in $\mathbb{R}^{4}$, namely

$K=\left\{ \left( 
\begin{array}{c}
a \\ 
x \\ 
y \\ 
z%
\end{array}%
\right) \in \mathbb{R}^{4}:a\geq 0\text{ and }a^{2}\geq
x^{2}+y^{2}+z^{2}\right\} =$

$=\left\{ \left( 
\begin{array}{c}
a \\ 
v%
\end{array}%
\right) \in \mathbb{R}^{4}:a\geq 0,\text{ }v\in \mathbb{R}^{3}\text{ and }%
a^{2}\geq \left\Vert v\right\Vert ^{2}\right\} .$

\bigskip

The Stokes cone $K$ has the following properties:

\noindent \textbf{1.a.} For all $\left( 
\begin{array}{c}
a \\ 
v%
\end{array}%
\right) ,\left( 
\begin{array}{c}
b \\ 
w%
\end{array}%
\right) $ vectors in $K$ and $\alpha ,\beta \geq 0$, we have

$\alpha \left( 
\begin{array}{c}
a \\ 
v%
\end{array}%
\right) +\beta \left( 
\begin{array}{c}
b \\ 
w%
\end{array}%
\right) $ is in $K$.

\noindent \textbf{1,b}. $\left( 
\begin{array}{c}
0 \\ 
\overline{0}%
\end{array}%
\right) \in K$. And if $\left( 
\begin{array}{c}
a \\ 
v%
\end{array}%
\right) $ and $\left( 
\begin{array}{c}
-a \\ 
-v%
\end{array}%
\right) $ are both in $K$, then $\left( 
\begin{array}{c}
a \\ 
v%
\end{array}%
\right) =\left( 
\begin{array}{c}
0 \\ 
\overline{0}%
\end{array}%
\right) $, and

\noindent \textbf{1.c}. If there exists a limit of vectors in $K$, $%
\lim_{m\rightarrow \infty }\left( 
\begin{array}{c}
a_{m} \\ 
v_{m}%
\end{array}%
\right) =\left( 
\begin{array}{c}
\widetilde{a} \\ 
\widetilde{v}%
\end{array}%
\right) $, then $\left( 
\begin{array}{c}
\widetilde{a} \\ 
\widetilde{v}%
\end{array}%
\right) $ also belongs to $K$. So $K$ is a closed subset of $\mathbb{R}^{4}$%
, and its boundary, denoted by $\partial K$, is

\begin{center}
$\partial K=\left\{ \left( 
\begin{array}{c}
\alpha \\ 
v%
\end{array}%
\right) \in \mathbb{R}^{4}:v\in \mathbb{R}^{3}\text{, }\alpha \geq 0\text{
and }\left\Vert v\right\Vert ^{2}=\alpha ^{2}\right\} $.
\end{center}

\bigskip

Therefore, the interior of $K$ is a nonempty set. In this case, we say that $%
K$ is a solid cone, see the Appendix.

\bigskip

\noindent \textbf{1.d. }The set $\mathcal{C}=\left\{ \left( 
\begin{array}{c}
1 \\ 
v%
\end{array}%
\right) \in K:\left\Vert v\right\Vert =1\right\} $ is a \textbf{%
cross-section }of $K$. That is, for each nonzero vector $\left( 
\begin{array}{c}
\alpha \\ 
v%
\end{array}%
\right) \in K$, there exist uniques $\beta >0$ and $\left( 
\begin{array}{c}
1 \\ 
y%
\end{array}%
\right) \in \mathcal{C}$ such that $\beta \left( 
\begin{array}{c}
1 \\ 
y%
\end{array}%
\right) =\left( 
\begin{array}{c}
\alpha \\ 
v%
\end{array}%
\right) $.

The cross-sections are important in Cone theory because they give a way to
study cones by means of convex compact sets.

\bigskip

\section{\noindent \textbf{The Mueller matrices}}

\noindent \textbf{Notation}. We will describe the $4\times 4-$real matrices
as $M=\left[ 
\begin{array}{cc}
1 & w_{0}^{T} \\ 
v_{0} & m%
\end{array}%
\right] $ where $v_{0},w_{0}\in \mathbb{R}^{3}$, $m\in M_{3}(\mathbb{R})$ be
a $3\times 3-$real matrix and $X^{T}$ denotes the transpose of a matrix or
vector $X.$

A matrix $M$ is a \textbf{Mueller matrix} if as linear function, $M$ sends
Stokes vectors to Stokes vectors. In other words, $M$ leaves the Stokes cone
invariant, $M(K)\subseteq K$, and since the set of matrices that leave a
solid cone invariant is a solid cone, the set of the Mueller matrices forms
a solid cone in $M_{4}(\mathbb{R})$, which we will call the \textbf{Mueller
cone}

\begin{center}
$\widetilde{K}=\{M\in M_{4}(\mathbb{R}):M(K)\subseteq K\}$
\end{center}

\noindent with the matrix norm $\left\Vert M\right\Vert _{2}=\sup \left\{
\left\Vert M\left( 
\begin{array}{c}
\alpha \\ 
v%
\end{array}%
\right) \right\Vert :\left\Vert \left( 
\begin{array}{c}
\alpha \\ 
v%
\end{array}%
\right) \right\Vert =1\right\} $

\begin{center}
$\underset{\text{finite dim}}{=}\max \left\{ \left\Vert M\left( 
\begin{array}{c}
\alpha \\ 
v%
\end{array}%
\right) \right\Vert :\left\Vert \left( 
\begin{array}{c}
\alpha \\ 
v%
\end{array}%
\right) \right\Vert =1\right\} $,
\end{center}

\noindent see (4.5.5).

Note that since all the norms in a finite dimension real vector space are
equivalent, for technical reasons, we chose the matrix norm $\left\Vert
-\right\Vert _{2}$.

These and the next properties make the Mueller cone a powerful tool to study
both Stokes vectors and Mueller matrices.

\bigskip

\subsection{\noindent \textbf{Properties of the Mueller Cone}}

\bigskip

\noindent \textbf{2.1.0}. Take the following set of Stokes vectors

$\mathcal{C}=\left\{ \left( 
\begin{array}{c}
1 \\ 
v%
\end{array}%
\right) \in \mathbb{R}^{4}:\left\Vert v\right\Vert =1\right\} .$

Using that $K$ is a cone, to prove that a matrix $M$ is Mueller, it is
enough to prove that $M($ $\mathcal{C})\subseteq K$, that is, $M$ sends each
vector of $\mathcal{C}$\ to $K$.

\noindent \textbf{2.1.1.} The following are Mueller matrices: The identity $%
I_{4}=\left[ 
\begin{array}{cc}
1 & \overline{0}^{T} \\ 
\overline{0} & I_{3}%
\end{array}%
\right] $, the zero matrix $\left[ 
\begin{array}{cc}
0 & \overline{0}^{T} \\ 
\overline{0} & [0]%
\end{array}%
\right] $, $G=\left[ 
\begin{array}{cc}
1 & \overline{0}^{T} \\ 
\overline{0} & -I_{3}%
\end{array}%
\right] $, $E_{11}=\left[ 
\begin{array}{cc}
1 & \overline{0}^{T} \\ 
\overline{0} & [0]%
\end{array}%
\right] $, $E_{11}+E_{ij}$ where the $pq-$coordinate of $E_{ij}$ is

\bigskip

\begin{center}
$E_{ij}(p,q)=\left\{ 
\begin{array}{cc}
1 & \text{if }(p,q)=(i.j) \\ 
0 & \text{otherwise}%
\end{array}%
\right. $
\end{center}

\noindent \textbf{2.1.2}. Let $M=\left[ 
\begin{array}{cc}
a & w_{0}^{T} \\ 
v_{0} & m%
\end{array}%
\right] $ be a Mueller matrix. Then $\left( 
\begin{array}{c}
a \\ 
v_{0}%
\end{array}%
\right) ,\left( 
\begin{array}{c}
a \\ 
w_{0}%
\end{array}%
\right) $ are Stokes vectors.

\bigskip

\noindent \textbf{2.1.3}. Let $M=\left[ 
\begin{array}{cc}
a & w_{0}^{T} \\ 
v_{0} & m%
\end{array}%
\right] $ be a Mueller matrix with $a=0$. Then $M$ must be the zero matrix.

\bigskip

Therefore, if $a\neq 0$ we will use the \textbf{normalized matrix of }$M$, $%
\frac{1}{a}M$, which is also a Mueller matrix, since $\widetilde{K}$ is a
cone.

In the last section, we include a computer program to decide if a $4\times 4$%
-real matrix is Mueller, using van der Mee's quadratic form, see (2.1.5)
below.

\bigskip

\noindent \textbf{2.1.4.1. }Let $v_{0}\in \mathbb{R}^{3}$ be a vector with
norm $\left\Vert v_{0}\right\Vert \leq 1$ and $m\in M_{3}(\mathbb{R})$ be a $%
3\times 3-$matrix of norm $\left\Vert m\right\Vert _{2}\leq 1.$ Then

$\bigskip $

$\left[ 
\begin{array}{cc}
1 & v_{0}^{T} \\ 
\overline{0} & [0]%
\end{array}%
\right] ,\left[ 
\begin{array}{cc}
1 & \overline{0}^{T} \\ 
v_{0} & [0]%
\end{array}%
\right] ,\left[ 
\begin{array}{cc}
1 & \overline{0}^{T} \\ 
\overline{0} & m%
\end{array}%
\right] $ are Mueller matrices.

\bigskip

Moreover, if two nonzero vectors $v_{0},w_{0}$ in $\mathbb{R}^{3}$ satisfy $%
\left\Vert v_{0}\right\Vert +\left\Vert w_{0}\right\Vert =1$, the matrix $%
\left[ 
\begin{array}{cc}
1 & w_{0}^{T} \\ 
v_{0} & [0]%
\end{array}%
\right] $ is Mueller.

\bigskip

\noindent \textbf{2.1.4.2.} Let $\left[ 
\begin{array}{cc}
1 & w_{0}^{T} \\ 
v_{0} & m%
\end{array}%
\right] $ be a Mueller matrix. Using that $G=\left[ 
\begin{array}{cc}
1 & \overline{0}^{T} \\ 
\overline{0} & -I_{3}%
\end{array}%
\right] $ is Mueller and $\widetilde{K}$ is a cone, the following are
Mueller matrices:

$\left[ 
\begin{array}{cc}
1 & w_{0}^{T} \\ 
\overline{0} & [0]%
\end{array}%
\right] ,\left[ 
\begin{array}{cc}
1 & -w_{0}^{T} \\ 
\overline{0} & [0]%
\end{array}%
\right] ,\left[ 
\begin{array}{cc}
1 & \overline{0}^{T} \\ 
v_{0} & [0]%
\end{array}%
\right] ,$

$\left[ 
\begin{array}{cc}
1 & \overline{0}^{T} \\ 
-v_{0} & [0]%
\end{array}%
\right] ,\left[ 
\begin{array}{cc}
1 & \overline{0}^{T} \\ 
\overline{0} & m%
\end{array}%
\right] ,\left[ 
\begin{array}{cc}
1 & \overline{0}^{T} \\ 
\overline{0} & -m%
\end{array}%
\right] $

Therefore, if $\left[ 
\begin{array}{cc}
1 & w_{0}^{T} \\ 
v_{0} & m%
\end{array}%
\right] $ is Mueller, its norm satisfies $\left\Vert m\right\Vert _{2}\leq
1. $

\bigskip

The proofs of (2.1.4.2) are straightforward using that a matrix $M$ is
Mueller if and only if $M(\mathcal{C})\subseteq K$ where $\mathcal{C}%
=\left\{ \left( 
\begin{array}{c}
1 \\ 
v%
\end{array}%
\right) \in \mathbb{R}^{4}:\left\Vert v\right\Vert =1\right\} .$

\bigskip

\noindent \textbf{2.1.5. Van der Mee's quadratic form} [vderM93].

\bigskip

\noindent \textbf{2.1.5.1.} The quadratic form $q_{G}$\ characterizes the
Stokes vectors, in the following way:

\bigskip

\noindent $\left( 
\begin{array}{c}
a \\ 
z%
\end{array}%
\right) \in K$ if and only if\textbf{\ }$a\geq 0$ and $q_{G}\left( 
\begin{array}{c}
a \\ 
z%
\end{array}%
\right) \geq 0$, \noindent where $G=\left[ 
\begin{array}{cc}
1 & \overline{0}^{T} \\ 
\overline{0} & -I_{3}%
\end{array}%
\right] .$

\bigskip

If $a=0$ then $\left( 
\begin{array}{c}
a \\ 
z%
\end{array}%
\right) =\left( 
\begin{array}{c}
0 \\ 
\overline{0}%
\end{array}%
\right) .$

If $a>0$ we have the following

\noindent\ $q_{G}\left( 
\begin{array}{c}
a \\ 
z%
\end{array}%
\right) =\left( 
\begin{array}{cc}
a & z^{T}%
\end{array}%
\right) G\left( 
\begin{array}{c}
a \\ 
z%
\end{array}%
\right) $ is $\left\{ 
\begin{array}{cc}
>0 & \left( 
\begin{array}{c}
a \\ 
z%
\end{array}%
\right) \in int_{\mathbb{R}^{3}}(K)\cup int_{\mathbb{R}^{3}}(-K) \\ 
=0 & \left( 
\begin{array}{c}
a \\ 
z%
\end{array}%
\right) \in \partial (K)\text{ \ \ \ \ \ \ \ \ \ } \\ 
<0 & \text{elsewhere \ \ \ \ \ \ \ \ \ \ \ \ \ \ \ \ \ \ \ \ \ \ \ \ \ \ \ \ 
}%
\end{array}%
\right. $

\bigskip

\noindent \textbf{2.1.5.2.} Let $M$ be a $4\times 4-$real matrix and $q_{M}$
be the quadratic form defined by $M^{T}GM$. Then

\bigskip

$M$ is a Mueller matrix if and only if for all $\left( 
\begin{array}{c}
a \\ 
z%
\end{array}%
\right) \in K$ with $\left( 
\begin{array}{c}
b \\ 
x%
\end{array}%
\right) =M\left( 
\begin{array}{c}
a \\ 
z%
\end{array}%
\right) $, $b\geq 0$ and $q_{M}\left( 
\begin{array}{c}
b \\ 
x%
\end{array}%
\right) \geq 0$.

\bigskip

\noindent \textbf{2.1.5.3.} Observation: Solely $q_{M}$ does not
characterize Mueller matrices. For example,

$M=\left[ 
\begin{array}{cc}
-1 & \overline{0} \\ 
\overline{0} & [0]%
\end{array}%
\right] $ is not a Mueller matrix, but $q_{M}(K)\geq 0$.

Namely, $M^{T}GM=\left[ 
\begin{array}{cc}
-1 & \overline{0} \\ 
\overline{0} & [0]%
\end{array}%
\right] \left[ 
\begin{array}{cc}
1 & \overline{0} \\ 
\overline{0} & -I_{3}%
\end{array}%
\right] \left[ 
\begin{array}{cc}
-1 & \overline{0} \\ 
\overline{0} & [0]%
\end{array}%
\right] =\left[ 
\begin{array}{cc}
1 & \overline{0} \\ 
\overline{0} & [0]%
\end{array}%
\right] $

\noindent and for all $\left( 
\begin{array}{c}
a \\ 
z%
\end{array}%
\right) \in K$, $q_{M}\left( 
\begin{array}{c}
a \\ 
z%
\end{array}%
\right) =a^{2}\geq 0$.

\bigskip

\noindent \textbf{2.1.6. }By Sekera [S1966] or Ossikovski [02009], the
transpose of a Mueller matrix is Mueller.

\noindent \textbf{2.1.7.} By (4.2) and since the Mueller cone is solid, it
contains linear bases of the $4\times 4-$real matrices. For example, the
following set is a linear basis of $M_{4}(\mathbb{R})$ of Mueller matrices:

\begin{center}
$\widetilde{\mathcal{B}}=\{E_{11}+E_{ij}:1\leq i,j\leq 4\}$
\end{center}

\noindent where $E_{ij}\in M_{4}(\mathbb{R})$ is the matrix whose $ij$th
entry is $1$ and $0$ elsewhere.

\bigskip

\subsection{\noindent \textbf{Approximations of a matrix by Mueller matrices}%
}

There are several ways to approximate a matrix by means of Mueller matrices.
Here we explain some methods.

\noindent \textbf{2.2.1. Approximation of a matrix by Mueller matrices.}

The first approximation is using the fact that the matrix $2E_{11}=\left[ 
\begin{array}{cc}
2 & \overline{0}^{T} \\ 
\overline{0} & [0]%
\end{array}%
\right] $ is in the interior of the Mueller cone $\widetilde{K}$, and that
the closed ball of radius $1$ and center $2E_{11}$ is contained in the
interior of $\widetilde{K}$, see (4.5.5.4). Then for each nonzero matrix $%
M\in M_{4}(\mathbb{R})$ the matrices $M+2\left\Vert M\right\Vert _{2}E_{11}$
and $-M+2\left\Vert M\right\Vert _{2}E_{11}$\ are Mueller matrices.
Moreover, they are $K-$primitive matrices, see (4.5.5.3).

\bigskip

\noindent \textbf{2.2.2. Approximation of a Mueller matrix by invertible
Mueller matrices. }

\bigskip

\noindent \textbf{2.2.2.1. Main approximation. }Let $\varepsilon >0$, such
that $\varepsilon $ and $-\varepsilon $ are not eigenvalues of $M$. Then $%
\varepsilon I_{4}+M$ is an approximation of $M$ by an invertible matrix.

\bigskip

Namely, recall that the identity matrix is Mueller.

Let $M$ be a Mueller matrix. Then, since $\widetilde{K}$ is a cone, for any $%
\varepsilon >0$, $\varepsilon I_{4}+M$ is Mueller.

In other hand, $\lambda $ is an eigenvalue of $M$ if and only if $%
\varepsilon +\lambda $ is an eigenvalue of $\varepsilon I_{4}+M$.

Since $\varepsilon $ and $-\varepsilon $ are not eigenvalues of $M$, $%
\lambda +\varepsilon \neq 0$ and then $\varepsilon I_{4}+M$ is invertible.

Therefore, $\varepsilon I_{4}+M$ is an invertible Mueller matrix.

\bigskip

The following aproximations are special cases of the main one.

\noindent \textbf{2.2.2.2}. Let $A=\left[ 
\begin{array}{cc}
a & w_{0}^{T} \\ 
v_{0} & m%
\end{array}%
\right] \in M_{4}(\mathbb{R})$ be a matrix and $S=\{\left\Vert \lambda
\right\Vert :\lambda \neq 0$ an eigenvalue of $A\}.$ If $S=\emptyset $
define $\varepsilon =1/100$. If $S\neq \emptyset $ then define $\varepsilon
= $min$\{1/100,$min$\{S\}\}$. The matrix

\begin{center}
$\varepsilon I_{4}+(A+2\left\Vert A\right\Vert _{2}E_{11})$
\end{center}

\noindent is an invertible Mueller matrix.

This is the approximation that is used in the programs.

\bigskip

\noindent \textbf{2.2.2.3}. In general, for any nonzero invertible matrix $%
\left[ 
\begin{array}{cc}
a & w_{0}^{T} \\ 
v_{0} & m%
\end{array}%
\right] =A\in M_{4}(\mathbb{R})$, $A+2\left\Vert A\right\Vert _{2}E_{11}$ is
Mueller. Then for every $\varepsilon >0$, the matrix 
\begin{equation*}
(\rho _{(A+2\left\Vert A\right\Vert _{2}E_{11})}+\varepsilon
)I_{4}+(A+2\left\Vert A\right\Vert _{2}E_{11})
\end{equation*}
is Mueller and invertible, where $\rho _{(A+2\left\Vert A\right\Vert
_{2}E_{11})}$ is the spectral radius of $A+2\left\Vert A\right\Vert
_{2}E_{11}$, see (4.7).

In other hand, if $A$ is invertible, we have two cases:

$\cdot $ If $A+2\left\Vert A\right\Vert _{2}E_{11}$ is invertible (also is
Mueller).

$\cdot $ If $A+2\left\Vert A\right\Vert _{2}E_{11}$ is not invertible, we
have $0=\det (A+2\left\Vert A\right\Vert _{2}E_{11})=2\left\Vert
A\right\Vert _{2}\det m$ $+\det A$ and since $\det A\neq 0$, $\det m\neq 0$.
Therefore $A+3\left\Vert A\right\Vert _{2}E_{11}$ is Mueller and invertible.

\bigskip

\noindent \textbf{2.2.3}. \textbf{Approximation of a Mueller matrix by
Mueller }$K-$\textbf{primitive matrices.}

Let $M$ be a Mueller matrix, then for all $n\geq 1$, $M+\frac{2}{n}E_{11}$
is $K-$primitive since $M+\frac{2}{n}E_{11}$ is in the interior of $%
\widetilde{K}$, see (4.4.1), and\bigskip

\begin{center}
$M=\lim_{n\rightarrow \infty }M+\frac{2}{n}E_{11}$
\end{center}

\bigskip

\section{\noindent \textbf{An Application: A Computer Algorithm of a
Polimeter's Eigenvalues Calibration Method (ECM)}}

Based on the Eigenvalues Calibration Method (ECM) of Eric Compain, St\'{e}%
phane Poirier, and Bernard Drevillon [CPD99], we give a computational
program of this method.

Calibration of polarization-state generators PSG's, polarimeters, and
Mueller-matrix ellipsometers MME's is an important factor in the practical
use of these instruments. In the ECM, the PSG and the polarimeter are
described by two $4\times 4$ matrices $W$ and $A$, and their 32 coefficients
are determined from three or four measurements performed on reference
samples.

\bigskip

\textbf{The steps of our computacional program for the Eigenvalues
Calibration Method.}

\noindent \textbf{3.1}. We have three $4\times 4-$real\ matrices: a matrix $%
M $ the target calibration matrix, and two matrices resulting from observing
through the polarimeter, $aw$ without biological sample and $amw$ with
biological sample.

The ECM gives an invertible matrix $W$, as mentioned above, and the new
matrix $M$ is an approximation by an invertible Mueller matrix of $%
W(aw)^{-1}amwW^{-1}$.

\bigskip

\noindent \textbf{3.2}. Matrix $aw$ must be invertible, otherwise our
program approximates it to an invertible matrix, see (2.2.2.1).

\bigskip

\noindent \textbf{3.3}. To find the matrix $W$, the following linear
function is constructed

\begin{center}
$%
\begin{array}{cccc}
\mathbb{H}: & M_{4}(\mathbb{R}) & \longrightarrow & M_{4}(\mathbb{R})\text{
\ \ \ \ \ \ \ \ \ \ \ \ \ \ \ \ \ \ } \\ 
& X & \longmapsto & MX-X(aw)^{-1}(amw)%
\end{array}%
$
\end{center}

\noindent \textbf{3.4.} If $\ker \mathbb{H}\neq \{0\}$ then let's take any
invertible matrix $W$ in the kernel.

If there isn't one, take any non-zero matrix and go to (3.6).

\noindent \textbf{3.5.} If $\ker \mathbb{H}=\{0\}$ we have two cases:

\textbf{3.5.1} If $\mathbb{H}$ have real eigenvalues then take the smallest
eigenvalue {}{}of the matrix $\mathbb{H}$, its respective eigenvector $W$
and go to (3.6).

\textbf{3.5.2.} If all the eigenvalues of $\mathbb{H}$ are non-real numbers,
then we calculate the smallest eigenvalue of $\mathbb{H}^{T}\mathbb{H}$,
whose eigenvalues are all real, then take its respective eigenvector $W$ and
go to (3.6).

\noindent \textbf{3.6.} If the resulting $W$ is not invertible we
approximate it to an invertible matrix.

\noindent \textbf{3.7.}The new Mueller matrix $M$ will be the approximation
by an invertible Mueller matrix of $W(aw)^{-1}amwW^{-1}$ using the
approximation in 2.2.2.2.

\bigskip

\section{\noindent {\protect\LARGE Appendix. Basic notions of Theory of
Cones.}}

For this section we recommend: [N68], [B81], [BS75], [Bi67] and [V68].

Let $%
\mathbb{R}
^{n}$ be the real vector space of finite dimension $n$ with its usual norm.

\noindent \textbf{4.1.} A subset $K$ of\ $%
\mathbb{R}
^{n}$\ is a \textbf{cone }if it satisfies the following

\noindent a) $K$\ is closed $%
\mathbb{R}
^{n}$\ (with the topology given by its norm).

\noindent b) For all $x,y\in K$ and $\alpha ,\beta \geq 0,$\ $\alpha x+\beta
y\in K$.

\noindent c) $K\cap (-K)=\{\overline{0}\}$,\ where $-K=\{-x;$\ $x\in K\}$.

\bigskip

\noindent \textbf{4.2.}{\small \ }A cone $K$\ is a \textbf{solid cone in} $%
\mathbb{R}
^{n}$ if its interior $int_{\mathbb{R}^{n}}K$ is not empty (ie., there exist 
$z\in K,r>0$ such that the closed ball $B_{r}(z)$, of radius $r$ and center $%
z$, is contained in $K$).

If $z\in int_{\mathbb{R}^{n}}K$ then for all $\alpha >0$, $\alpha z\in int_{%
\mathbb{R}^{n}}K$.

$K$ is a solid cone in $%
\mathbb{R}
^{n}$ if and only if $K$ contains a basis of $%
\mathbb{R}
^{n}$.

Therefore, $K$ is a solid cone in its Span($K$) (the vector space generated
by $K$).

\bigskip

\noindent \textbf{4.3.} A compact (closed and bounded) subset $\mathcal{C}$
of $K$ is a \textbf{cross-section} if for every non-zero vector $v$ of $K$
there exist uniques $\alpha >0$ and $y\in \mathcal{C}$ such that $x=\alpha
y. $

\bigskip

\noindent \textbf{4.4}. Let $K$ be a solid cone in $%
\mathbb{R}
^{n}$. Then

\textbf{4.4.1}. Let $z$ be a vector of $K$. Then

$z\in int_{\mathbb{R}^{n}}K$ if and only if for all $x\in 
\mathbb{R}
^{n}$ there exists $\lambda >0$ with $z-\lambda x$ in $K$.

\textbf{4.4.2}. If $x,y\in K$ such that $y\in int_{\mathbb{R}^{n}}K$ and $%
x-y\in K$, then $x\in int_{\mathbb{R}^{n}}K$.

\textbf{4.4.3.} Let $y\neq \overline{0}$ and $\left\langle y,-\right\rangle :%
\mathbb{R}
^{n}\longrightarrow 
\mathbb{R}
$ be the usual inner product of $%
\mathbb{R}
^{n}$. Then the kernel $\ker \left\langle y,-\right\rangle $ is a hyperplane
which divides $%
\mathbb{R}
^{n}$ into two disjoint sets $\{v\in 
\mathbb{R}
^{n}:\left\langle y,v\right\rangle >0\}$ and $\{v\in 
\mathbb{R}
^{n}:\left\langle y,v\right\rangle \leq 0\}$. Moreover, if $K$ is a solid
cone then there exists $y\in int_{\mathbb{R}^{n}}K$ such that $K\subseteq
\{v\in 
\mathbb{R}
^{n}:\left\langle y,v\right\rangle >0\}$.

\noindent \textbf{4.5. Examples of cones}

\noindent \textbf{4.5.0. }The only vector space that is a cone is $\{%
\overline{0}\}$.

\noindent \textbf{4.5.1.} If $K$ is a cone, then $-K$ is also a cone.

\noindent \textbf{4.5.2.} $(%
\mathbb{R}
^{n})^{+}=\{(v_{1},...,v_{n})\in 
\mathbb{R}
^{n};$ $v_{i}\geq 0,$ for $i=1,...,n\}$ is a solid cone called the \textbf{%
positive }(or nonnegative) \textbf{cone} of\textit{\ }$%
\mathbb{R}
^{n}.$

\noindent \textbf{4.5.3. }The \textbf{ice cream cone} $\{(a,x,y)\in 
\mathbb{R}
^{3};$ $a\geq 0$ and$\ x^{2}+y^{2}\leq a^{2}\}$ is a solid cone of $%
\mathbb{R}
^{3}$.

\noindent \textbf{4.5.4.} The \textbf{Stokes cone} is the ice cream cone of $%
\mathbb{R}
^{4}$:

$K=\left\{ \left( 
\begin{array}{c}
a \\ 
v%
\end{array}%
\right) \in 
\mathbb{R}
^{4}:v\in 
\mathbb{R}
^{3}\text{, }a\geq 0\text{ and}\ \left\Vert v\right\Vert ^{2}\leq
a^{2}\right\} $.

$K$ is a solid cone, moreover the closed ball of radius $1$ and center the
vector $\left( 
\begin{array}{c}
2 \\ 
\overline{0}%
\end{array}%
\right) $, $B_{1}\left( \left( 
\begin{array}{c}
2 \\ 
\overline{0}%
\end{array}%
\right) \right) \subseteq int_{\mathbb{R}^{4}}K$ the interior of $K$.

$\mathcal{C}=\left\{ \left( 
\begin{array}{c}
1 \\ 
v%
\end{array}%
\right) \in 
\mathbb{R}
^{4}:v\in 
\mathbb{R}
^{3}\text{ and}\ \left\Vert v\right\Vert =1\right\} $ is a cross-section of $%
K$.

\bigskip

\noindent In order to figure out properties of the Stokes cone, it is good
to see them first in the ice cream cone of $%
\mathbb{R}
^{3}$.

\bigskip

\noindent \textbf{4.5.5.\ The set of matrices which leave a solid cone
invariant is a solid cone.}

Let $K$ be solid cone of $%
\mathbb{R}
^{n}$. Then $\widetilde{K}=\{A\in M_{n}(\mathbb{R}):AK\subseteq K\}$ is a
solid cone in the space vector of the $n\times n-$real matrices, $M_{n}(%
\mathbb{R})$. And since any two norms are equivalent, we choose the
matricial norm

$\left\Vert A\right\Vert _{2}=\sup \{\left\Vert Av\right\Vert :\left\Vert
v\right\Vert =1$ with $v\in 
\mathbb{R}
^{n}\}\underset{\text{finite dim}}{=}\max \{\left\Vert Av\right\Vert
:\left\Vert v\right\Vert =1$ with $v\in 
\mathbb{R}
^{n}\}$.

\bigskip

It is easy to prove that $\widetilde{K}$ is a cone. Now we will prove that $%
\widetilde{K}$ is solid.

\noindent \textbf{4.5.5.1}. For all $A,M\in \widetilde{K}$ we have that $%
A-M\in \widetilde{K}$ if and only if for each $v\in K$, $Av-Mv\in K$.

\noindent \textbf{4.5.5.2. }The interior of $\widetilde{K}$, $int_{M_{n}(%
\mathbb{R})}(\widetilde{K})$ is the set $\{A\in M_{n}(\mathbb{R}):$ $A(K-\{%
\overline{0}\})\subseteq int_{\mathbb{R}^{n}}K\}$.

\noindent Sketch of the proof of (4.5.5.2): Since $K$ is a solid cone, by
(4.4.3), there exists $z\in int_{\mathbb{R}^{n}}K$ such that $\left\Vert
z\right\Vert =1$ and $K\subseteq \{x\in \mathbb{R}^{4}:\left\langle
z,x\right\rangle >0\}$. Now define the linear function\bigskip

$%
\begin{array}{cccc}
M_{z}: & \mathbb{R}^{n} & \longrightarrow & \mathbb{R}^{n} \\ 
& x & \longmapsto & \left\langle z,x\right\rangle z%
\end{array}%
$ with associated matrix $M_{z}=\left[ 
\begin{array}{ccc}
z_{1}^{2} & \cdots & z_{n}z_{1} \\ 
z_{1}z_{2} &  & z_{n}z_{2} \\ 
\vdots & \ddots & \vdots \\ 
z_{1}z_{n} & \cdots & z_{n}^{2}%
\end{array}%
\right] $. Observe that $\left\Vert M_{z}\right\Vert _{2}=1$.

\bigskip

Take $A\in int_{M_{n}(\mathbb{R})}(\widetilde{K})$, then there exists $r>0$
such that $B_{r}(A)\subseteq \widetilde{K}$, then $A-rM_{z}\in \widetilde{K}$%
.

Therefore, for all $v\in K-\{\overline{0}\}$, $Av-r\left\langle
z.v\right\rangle z\in K$, and since $r\left\langle z.v\right\rangle z\in
int_{\mathbb{R}^{n}}K$, by (4.4.1), $Av\in int_{\mathbb{R}^{n}}K$.

Therefore, $A(K-\{\overline{0}\})\subseteq int_{\mathbb{R}^{n}}K$.

\bigskip

To prove the other inclusion: Let $A$ be a matrix such that $A(K-\{\overline{%
0}\})\subseteq int_{\mathbb{R}^{n}}K$.\bigskip

By (4.4.1), it is enough to prove that for all nonzero matrix $M$, there
exists $\lambda >0$ such that $A-\lambda M\in \widetilde{K}.$

\noindent $\bullet $ Since $A(K-\{\overline{0}\})\subseteq int_{\mathbb{R}%
^{n}}K$. For each nonzero vector $v\in K$, $\frac{Av}{\left\Vert
Av\right\Vert }\in int_{\mathbb{R}^{n}}K$. Then there exists $r_{v}>0$ such
that the open ball $B_{r_{v}}^{0}(\frac{Av}{\left\Vert Av\right\Vert }%
)\subseteq K$.

\noindent $\bullet $ Now take the cross-section $\mathcal{C}=\{v\in
K:\left\Vert v\right\Vert =1\}$ of $K$, see (4.3). And since $A$ is
continous, $A\mathcal{C}$ is a compact set and $A\mathcal{C}\subseteq
\bigcup\limits_{v\in \mathcal{C}}B_{r_{v}}^{0}(\frac{Av}{\left\Vert
Av\right\Vert })$, then there exist $v_{1},...,v_{m}\in \mathcal{C}$ such
that $A\mathcal{C}\subseteq \bigcup\limits_{j=1}^{m}B_{r_{v_{j}}}^{0}(\frac{%
Av_{j}}{\left\Vert Av_{j}\right\Vert })\subseteq K$.

Denote by $\widetilde{r}:=\min \{r_{v_{j}}:j=1,...,m\}$ and by $v_{h}$ the
vector such that $\widetilde{r}=r_{v_{h}}$.

\noindent $\bullet $ To prove that for every nonzero matrix $M\in \widetilde{%
K}$, there exists $\lambda >0$, $A-\lambda M\in \widetilde{K}$.

Using that $A(K-\{\overline{0}\})\subseteq int_{\mathbb{R}^{n}}K$ and $%
\widetilde{r}=r_{v_{h}}$, we have that for all nonzero $v\in \mathcal{C}$,
there exists $\lambda >0$ such that $\frac{Av}{\left\Vert A\right\Vert _{2}}%
-\lambda \frac{Av_{h}}{\left\Vert Av_{h}\right\Vert }$ and $\frac{Av_{h}}{%
\left\Vert Av_{h}\right\Vert }-\widetilde{r}\frac{Mv}{\left\Vert
M\right\Vert _{2}}$ are vectors in $K$.

Then $\frac{Av}{\left\Vert A\right\Vert _{2}}-\lambda \frac{Av_{h}}{%
\left\Vert Av_{h}\right\Vert }+\lambda \frac{Av_{h}}{\left\Vert
Av_{h}\right\Vert }-\lambda \widetilde{r}\frac{Mv}{\left\Vert M\right\Vert
_{2}}\in K$\ .

Then for all $v\in \mathcal{C}$, $Av-\frac{\lambda \widetilde{r}}{\left\Vert
A\right\Vert _{2}\left\Vert M\right\Vert _{2}}Mv\in K$ and $A-$ $\frac{%
\lambda \widetilde{r}}{\left\Vert A\right\Vert _{2}\left\Vert M\right\Vert
_{2}}M\in \widetilde{K}$.

Therefore by (4.5.5.1), $A\in int_{M_{4}(\mathbb{R})}(\widetilde{K}%
)._{\blacksquare }$

\bigskip

\noindent Corollary. $Mz$ is in the interior of $\widetilde{K}$, when $z$ is
in the interior of $K$ and $K\subseteq \{v\in \mathbb{R}^{n}:\left\langle
z,v\right\rangle >0\}$.

\noindent \textbf{4.5.5.3}. Therefore, the interior of $\widetilde{K}$ is
the set of the $K-$primitive matrices, see (4.7).

\noindent \textbf{4.5.5.4. Example.} The Stokes cone $K$ is solid, then by
(4.5.5), the Mueller matrices form a solid cone, $\widetilde{K}$.

Moreover, denote by $E_{11}$ the Mueller matrix, whose $ij$th entry is $1$
if $(i,j)=(1,1)$ and $0$ otherwise.

It is useful to know that $E_{11}$ is in the interior of $\widetilde{K}$.
Moreover, by using that the closed ball $B_{1}\left( \left( 
\begin{array}{c}
2 \\ 
\overline{0}%
\end{array}%
\right) \right) \subseteq int_{\mathbb{R}^{4}}K$, one can prove that $%
B_{1}(2E_{11})\subseteq int_{M_{4}(\mathbb{R})}(\widetilde{K})$.

\bigskip

\noindent \textbf{4.6. The Faces of a Cone.}

\textit{"Cones" is to "Vector spaces" as "Faces" is to "(some but
sufficient) Vector subspaces"}

\bigskip

A subcone $F$ of a cone $K$ is a \textbf{face} of $K$ if for every pair $%
v,w\in K$ such that $v+w\in F$ then both $v,w\in F$, see Figure 1.

\bigskip

\begin{figure}[]
\centering
\includegraphics[width=.6\textwidth]{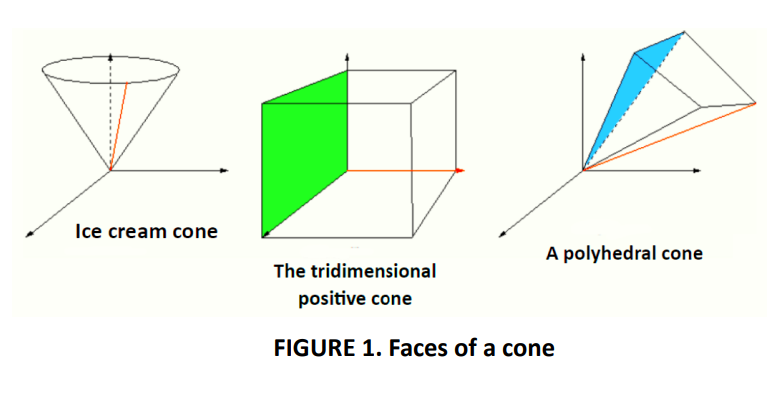}   
\caption{Faces of a cone.}
\label{fig:caras-cono-vw}
\end{figure}

\noindent \textbf{Examples.} \textbf{4.6.1.} $\{\overline{0}\}$ and $K$ are
always faces of $K$.

\textbf{4.6.2.} Every face different from $K$ is contained in the boundary
of $K.$

\textbf{4.6.3.} The ice cream cone of $%
\mathbb{R}
^{3}$ has an infinite number of faces (every semiray in the boundary is a
face), see (4.5.3).

\textbf{4.6.4.} A cone with at most a finite number of faces is called 
\textbf{polyhedral}. The positive cone of $%
\mathbb{R}
^{n}$ is polyhedral.

\bigskip

The next theorem is a beautiful generalization of the well known
Perron-Frobenius Theorem [HJ90, HJ99].

\noindent \textbf{4.7. THE BIRKHOFF-VANDERGRAFT'S THEOREM}

\bigskip

Let $A$ be a real $n\times n-$matrix and $K$ be a solid cone of $%
\mathbb{R}
^{n}$. We say that $A$ leaves $K$ invariant if for every vector $v$ in $K$, $%
Av$ is again in $K$ and denoted by $A(K)\subseteq K.$

\noindent Assume that $A$ leaves $K$ invariant. Then we say that

\noindent $A$ is $K-$\textbf{primitive} if $A(K-\{\overline{0}\})$ is
contained in the interior of $K$.

\noindent $A$ is $K-$\textbf{strongly irreducible} if there exists a number $%
m\geq 1$ such that $A^{m}(K-\{\overline{0}\})$ is contained in the interior
of $K$.

\noindent \noindent $A$ is $K-$\textbf{irreducible} If the only faces of $K$
that $A$ leaves invariant are $\left\{ \overline{0}\right\} $ and $K$.

\bigskip

\noindent Observations: $K-$primitive $\Rightarrow K-$strongly irreducible $%
\Rightarrow K-$irreducible, and

$A$ is $K-$irreducible if and only if $I+A$ is $K-$strongly irreducible.

\noindent

\noindent \textbf{Example. }Following example (4.5.3), the $\theta -$%
rotation over the $a-$axes with angle $0<\theta <2\pi $ of $K,$\ the ice
cream cone in $\mathbb{R}^{3},$ is $K-$irrreducible, but it is not $K-$%
strongly irreducible.

\bigskip

The \textbf{spectral radius} of a square matrix $A$ is denoted by $\rho _{A}$
and is the maximum among the modulus of the eigenvalues of $A$,

\bigskip

\begin{center}
$\rho _{A}=\max \{\left\Vert \lambda \right\Vert :\lambda \in \mathbb{C}$ an
eigenvalue of $A\}$
\end{center}

\bigskip

Observe that the spectral radius is not necessarily an eigenvalue. For
example, $\left[ 
\begin{array}{cc}
-2 & 0 \\ 
0 & 1%
\end{array}%
\right] $\bigskip\ has spectral radius $2$ and it is not an eigenvalue.

\noindent \textbf{4.7.1. Theorem(Birkhoff, 1967)}. Let $K$ be a solid cone
and a $A$ be a matrix which leaves $K$ invariant, then

(i) $\rho _{A}$ is an eigenvalue of $A$,

(ii) the degree of $\rho _{A}$ is no smaller than the degree of any other
eigenvalue having the same modulus,

(iii) $K$ contains an eigenvector corresponding to $\rho _{A}.$

Furthermore, conditions (i) and (ii) are sufficient to insure that $A$
leaves invariant a solid cone.

Recall that the \textbf{degree} of an eigenvalue $\lambda $\ of $A$ is the
size of the largest block of $\lambda $, in the Jordan canonical form of $A$.

\bigskip

\noindent \textbf{4.7.2. Proposition}. Let $K$ be a solid cone and a $A$ be
a non-zero matrix which leaves $K$ invariant, then

\noindent \textbf{(1)} $A$ is $K-$irreducible if and only if $A$ satisfies:
(i) $\rho _{A}$ is a simple eigenvalue of $A$, (ii) Any other eigenvalue
with modulus $\rho _{A}$ is simple, (iii) $K$ contains an eigenvector
corresponding to $\rho _{A}$ and it is the unique (up to scalar multiples)
eigenvector of $A$ belonging to $K.$

\noindent \textbf{(2)} $A$ is $K-$primitive if and only if $A$ is $K-$%
irreducible and $\rho _{A}$ is the unique eigenvalue having its modulus.

\bigskip

\noindent \textbf{4.7.3. Theorem(Birkhoff-Vandergraft, 1967-68). }Let $A$ be
a real $n\times n-$matrix which leaves a solid cone $K$ invariant. Assume $A$
is $K-$irreducible. Then

\bigskip

\noindent \textbf{a)} The spectral radius of $A$ is positive $\rho _{A}>0$
and is an eigenvalue of $A$.

\noindent \textbf{b)} There exists a unique (up to scalar multiples)
eigenvector $v$ in the interior of $K$, corresponding to $\rho _{A}$.

\noindent \textbf{c)} Moreover, if $A$ is $K-$strongly irreducible then for
each non-zero vector $w$ in $K$ the following limit exists

\begin{center}
$\lim_{m\rightarrow \infty }\frac{A^{m}w}{\rho _{A}^{m}}=\lambda _{w}v$ with 
$\lambda _{w}>0.$
\end{center}

\bigskip

\section{\noindent \textbf{COMPUTATIONAL PROGRAMS}}

In the following link are the computer programs of the article and others,
with detailed explanations:

https://github.com/IvanLopezR22/cone-of-mueller-matrices/tree/master

5.1. Matrix norm.

5.2. Van der Mee's quadratic form.

5.3. Know if the matrix is K-irreducible.

5.4. Know if the matrix is K-primitive.

5.5. Approximation by an invertible matrix.

5.6. Approximation by a Mueller matrix.

5.7. Approximation by an invertible Mueller matrix.

5.8. Approximation by a K-primitive matrix.

5.9. An Eigenvalue Calibration Method, [CPD99].

\bigskip

\noindent \textbf{Below we give the computer program of the ECM}:

The program was made using Python 3.11 and the libraries:

1. SymPy version 1.12

2. NumPy version 1.25.2

3. Matplotlib version 3.7.2

For a matrix $M = \left[ 
\begin{array}{cc}
a & w_{0}^{T} \\ 
v_{0} & m%
\end{array}%
\right] \in M_{4}(\mathbb{R})$, in the program we denote $M(inv)$, $M(mu)$
and $M(mu-inv)$ to the approximations by an invertible matrix, by a Mueller
matrix and by an invertible Mueller matrix respectively. 
\begin{lstlisting}[language=Python, title=Eigenvalues Calibration Method ]
from sympy import *
from sympy import Abs
import numpy as np
from numpy.linalg import eig
import sympy as sym
import matplotlib.pyplot as plt
import warnings

warnings.filterwarnings("ignore", category=RuntimeWarning)

np.set_printoptions(precision=15, suppress=True, linewidth=2000)


def request_matrix(matrix_name):
    print(f"Enter the values of the matrix {matrix_name}:")
    matrix_m = []
    for raw_position in range(4):
        raw = []
        for element in range(4):
            while True:
                try:
                    value = Rational(str(float((input(f"Enter the element {raw_position + 1}, {element + 1}: ")))))
                    if raw_position == 0 and element == 0 and value == 0:
                        print("The value at coordinate (1, 1) must be non-zero. Please try again.")
                        continue

                    raw.append(value)
                except ValueError:
                    print("A non-numeric character was entered, please input a decimal real number.")
                    continue

                else:
                    break

        matrix_m.append(raw)

    return Matrix(matrix_m)


def matrix_norm(main_matrix):
    mult_transpose_main = main_matrix.T * main_matrix
    eigenvalues_h, eigenvectors_h = eig(np.array(mult_transpose_main).astype(np.float64))
    norm_eigenvalues_mtm = []
    for i in range(len(eigenvalues_h)):
        norm_eigenvalues_mtm.append(Abs(N(re(eigenvalues_h[i]), 10)))
    s = sqrt(max(norm_eigenvalues_mtm))
    return s


def van_der_mee_theorem(main_matrix):
    G = Matrix([[1, 0, 0, 0], [0, -1, 0, 0], [0, 0, -1, 0], [0, 0, 0, -1]])
    x, y = sym.symbols('x,y')
    s = main_matrix * Matrix([1, x, y, sqrt(1 - x ** 2 - y ** 2)])
    t = G * (main_matrix * Matrix([1, x, y, sqrt(1 - x ** 2 - y ** 2)]))
    qm = s.dot(t)
    proy1_m = (main_matrix * Matrix([1, x, y, sqrt(1 - x ** 2 - y ** 2)]))[0]
    z_lamb_1 = sym.lambdify((x, y), qm)
    z_lamb_2 = sym.lambdify((x, y), proy1_m)
    a = 1
    b = 1
    xdata = np.linspace(-(b + .25), (b + .25), 1001)
    ydata = np.linspace(-(b + .25), (b + .25), 1001)
    X, Y = np.meshgrid(xdata, ydata)
    fig = plt.figure(figsize=(3.5 * 3.13, 1.5 * 3.13))
    R = (X ** 2 + Y ** 2 > a ** 2)
    expr_1 = sympify(qm)
    symbols_qm = expr_1.free_symbols
    expr_2 = sympify(proy1_m)
    symbols_m_1 = expr_2.free_symbols
    if not symbols_qm:
        K_1 = np.around(np.full((1001, 1001), qm).astype(np.float64), decimals=12)
    else:
        K_1 = np.around(z_lamb_1(X, Y), decimals=12)

    if not symbols_m_1:
        K_2 = np.around(np.full((1001, 1001), proy1_m).astype(np.float64), decimals=12)
    else:
        K_2 = np.around(z_lamb_2(X, Y), decimals=12)
    z_masked_1 = np.ma.masked_where(R, K_1)
    z_masked_2 = np.ma.masked_where(R, K_2)
    xmin_qm, ymin_qm = np.unravel_index(np.argmin(z_masked_1), z_masked_1.shape)
    mi_qm = (X[xmin_qm, ymin_qm], Y[xmin_qm, ymin_qm], z_masked_1.min())
    minimum_qm = z_masked_1.min()
    xmin_proy1_m, ymin_proy1_m = np.unravel_index(np.argmin(z_masked_2), z_masked_2.shape)
    mi_proy1_m = (X[xmin_proy1_m, ymin_proy1_m], Y[xmin_proy1_m, ymin_proy1_m], z_masked_2.min())
    minimum_proy1_m = z_masked_2.min()
    return qm, mi_qm, minimum_qm, proy1_m, mi_proy1_m, minimum_proy1_m, fig


def appr_mueller_matrix(w, m):
    e_1 = Matrix([[1, 0, 0, 0], [0, 0, 0, 0], [0, 0, 0, 0], [0, 0, 0, 0]])
    norm_m = matrix_norm(m)
    qm, mi_qm, minimum_qm, proy1_m, mi_proy1_m, minimum_proy1_m, fig = van_der_mee_theorem(m)
    if minimum_qm >= 0 and minimum_proy1_m >= 0:
        m_appr = m
        print(f"The minimum of the functions q{w} and proy1({w}) of the matrix {w} are "
              f"{minimum_qm} and {minimum_proy1_m} respectively.")
        print(f"Therefore, the matrix {w} is a Mueller matrix. Then {w} need not to be "
              f"approximated by a Mueller matrix.")
        name = str(w)
        return m_appr, name

    else:
        m_appr = ((2 * norm_m) * e_1) + m
        qm_1, mi_qm_1, minimum_qm_1, proy1_m_1, mi_proy1_m_1, minimum_proy1_m_1, fig_1 \
            = van_der_mee_theorem(m_appr)
        print(f"The minimum of the functions q{w} and proy1({w}) of the matrix {w} are "
              f"{minimum_qm} and {minimum_proy1_m} respectively.")
        print(f"Therefore, the matrix {w} is not a Mueller matrix. Then we approximate {w} to {w}(mu).")

        name = str(w) + "(mu)"

        return m_appr, name


def make_invertible(name, main_matrix):
    ident = Matrix([[1, 0, 0, 0], [0, 1, 0, 0], [0, 0, 1, 0], [0, 0, 0, 1]])
    if round(main_matrix.det(), 12) != 0:
        invertible_main_matrix = main_matrix
        print(f"The determinant of the matrix {name} is {round(main_matrix.det(), 12)}. Therefore {name} "
              f"is an invertible matrix. ")
    else:
        eigen_m = main_matrix.eigenvects()
        eigenvalues = []
        for element in eigen_m:
            eigenvalues.append((element[0]))
        norm_eigenvalues_m = []
        for i in range(len(eigen_m)):
            norm_eigenvalues_m.append(round(Abs(N(eigen_m[i][0], 12)), 12))
        norms_no_zero = [x for x in norm_eigenvalues_m if x != 0]
        if not norms_no_zero:
            norms_no_zero.append(2)
        eps = min([1 / 100, (1 / 2) * min(norms_no_zero)])
        invertible_main_matrix = (eps * ident) + main_matrix
        print(f"The determinant of the matrix {name} is {round(main_matrix.det(), 12)}. "
              f"Therefore {name} is not an invertible matrix. ")
        print("------------------------------------------------------------------------------------")
        
    return invertible_main_matrix


def appr_invert_mueller_matrix(w, main_matrix):
    m_mue_appr, name = appr_mueller_matrix(w, main_matrix)
    m_mue_inv = make_invertible(name, m_mue_appr)
    qm, mi_qm, minimum_qm, proy1_m, mi_proy1_m, minimum_proy1_m, fig = van_der_mee_theorem(m_mue_inv)
    print(f"Due to the previous data, the approximation {w}(mu-inv) of {w} by an invertible Mueller matrix is: "
          f"\n{np.array(m_mue_inv).astype(np.float64)}")
    print(f"The minimum of the functions q{w}(mu-inv) and proy1({w}(mu-inv)) of {w}(mu-inv) are "
          f"{minimum_qm} and {minimum_proy1_m} respectively.")
    print(f"The determinant of {w}(mu-inv) is {m_mue_inv.det()}. Therefore, {w}(mu-inv) is "
          f"an invertible Mueller matrix.")
    return m_mue_inv


def ecm(m):
    aw = make_invertible("aw", request_matrix("aw"))
    print(f"Then, the matrix aw is: \n{np.array(aw).astype(np.float64)}")
    print("------------------------------------------------------------------------------------")
    amw = request_matrix("amw")
    print('The matrix amw is: \n', np.array(amw).astype(np.float64))
    canonical_base = []
    for i in range(4):
        for j in range(4):
            def f(s, t):
                if s == i and t == j:
                    return 1
                else:
                    return 0

            k = Matrix(4, 4, f)
            canonical_base.append(k)

    aw_inv = aw.inv()
    image_base = []
    for element in canonical_base:
        w = (m * element) - (element * aw_inv * amw)
        image_base.append(w)
    h_matrix_columns = []
    for x in range(4):
        for y in range(4):
            row = []
            for element in image_base:
                row.append(element[x, y])
            h_matrix_columns.append(row)
    matrix_h = Matrix(h_matrix_columns)
    matrix_h_np = np.array(matrix_h).astype(np.float64)
    print(f"The matrix form of the function H in canonical basis is: \n{np.around(matrix_h_np, decimals=5)}")
    print("------------------------------------------------------------------------------------")
    null_space_h = matrix_h.nullspace()
    if null_space_h:
        print('The null space of H is non-trivial.')
        j = 0
        while j <= (len(null_space_h) - 1) and Abs(
                round(Matrix(np.array(null_space_h[j].transpose()).astype(np.float64).reshape((4, 4))).det(),
                      15)) == 0:
            j += 1
        else:
            if j == len(null_space_h):
                a = Matrix(np.array(null_space_h[0].transpose()).astype(
                    np.float64).reshape((4, 4)))
                print(f"No invertible matrices found in the nullspace of H, so we select the matrix: W=\n"
                      f"{np.array(a).astype(np.float64)}")
                print("------------------------------------------------------------------------------------")
            else:
                a = Matrix(np.array(null_space_h[j].transpose()).astype(
                    np.float64).reshape((4, 4)))
                print(f"An invertible matrix was found in the null space of H: W=\n{np.array(a).astype(np.float64)}")
                print("------------------------------------------------------------------------------------")
        w = make_invertible("W", Matrix(a))
        print(f"Then, the matrix W is: \n{np.array(w).astype(np.float64)}")
        print("------------------------------------------------------------------------------------")

    else:
        print('The null space of H is trivial.')
        eigenvalues_h, eigenvectors_h = eig(matrix_h_np)
        eigenvalues_h_without_error = []
        for i in eigenvalues_h:
            conj = i.conjugate()
            z = conj in eigenvalues_h
            if simplify(i).is_real == True:
                eigenvalues_h_without_error.append(i)
            elif simplify(i).is_real == False and z == False:
                eigenvalues_h_without_error.append(re(i))
            else:
                eigenvalues_h_without_error.append(i)
        print(f"The eigenvalues of H are: {eigenvalues_h_without_error}.")
        print("------------------------------------------------------------------------------------")
        eigenvectors_h_transpose = eigenvectors_h.transpose()
        real_eigenvalues = []
        complex_eigenvalues = []
        for i in eigenvalues_h_without_error:
            if simplify(i).is_real == True:
                real_eigenvalues.append(i)
            else:
                complex_eigenvalues.append(i)
        print(f"The real eigenvalues of H are: {real_eigenvalues}.")
        print(f"The complex eigenvalues of H are: {complex_eigenvalues}.")
        print("------------------------------------------------------------------------------------")
        if not real_eigenvalues:
            print('There are no real eigenvalues, therefore the approximation will be done in the matrix H^T*H.')
            ht_h = np.matmul(matrix_h_np.transpose(), matrix_h_np)
            eigenvalues_ht_h, eigenvectors_ht_h = eig(ht_h)
            eigenvectors_ht_h_transpose = eigenvectors_ht_h.transpose()
            print(f"The matrix H^{{T}}*H is: \n{ht_h}")
            print("------------------------------------------------------------------------------------")
            eigenvalues_ht_h_no_error = []
            for i in eigenvalues_ht_h:
                eigenvalues_ht_h_no_error.append(re(i))
            print(f"The eigenvalues of H^{{T}}*H are: {eigenvalues_ht_h_no_error}")
            print(f"The smallest eigenvalue of H^{{T}}*H is: {min(eigenvalues_ht_h_no_error)}")
            print("------------------------------------------------------------------------------------")
            a = np.array(
                eigenvectors_ht_h_transpose[eigenvalues_ht_h_no_error.index(min(eigenvalues_ht_h_no_error))]).astype(
                np.float64).reshape((4, 4))
            print(f"An eigenvector W associated to the eigenvalue {min(eigenvalues_ht_h_no_error)} is: \n{a}")
            print("------------------------------------------------------------------------------------")
            w = make_invertible("W", Matrix(a))
            print(f"Then, the matrix W is: \n{np.array(w).astype(np.float64)}")
            print("------------------------------------------------------------------------------------")
        else:
            abs_real_eigenvalues = []
            for i in range(len(real_eigenvalues)):
                abs_real_eigenvalues.append(Abs(real_eigenvalues[i]))
            print(f"The norms of the real eigenvalues of H are: {abs_real_eigenvalues}")
            print(f"The minimum of the norms of the eigenvalues is: {min(abs_real_eigenvalues)}")
            print("------------------------------------------------------------------------------------")
            ind_min_real_eigenvalues = eigenvalues_h_without_error.index(
                real_eigenvalues[abs_real_eigenvalues.index(min(abs_real_eigenvalues))])
            print(f"An eigenvalue associated with the minimum norm is:"
                  f"\n{eigenvalues_h_without_error[ind_min_real_eigenvalues]}")
            a = np.array(eigenvectors_h_transpose[ind_min_real_eigenvalues]).reshape((4, 4))
            print(f"An eigenvector W associated to the eigenvalue "
                  f"{eigenvalues_h_without_error[ind_min_real_eigenvalues]} is:\n{a}")
            print("------------------------------------------------------------------------------------")
            w = make_invertible("W", Matrix(a))
            print(f"Then, the matrix W is: \n{np.array(w).astype(np.float64)}")
            print("------------------------------------------------------------------------------------")
    new_m = w * aw_inv * amw * (w.inv())
    print()
    print(f"The new matrix M, denoted by New_M, is the matrix defined by new_M=W*(aw^{-1})*amw*(W^{-1})="
          f"\n{np.array(new_m).astype(np.float64)}")
    print("------------------------------------------------------------------------------------")
    print(f"Following the notation of the comments in the appr_invertible_mueller_matrix and "
          f"appr_by_mueller_matrix files, we call New_M(mu-inv) to the approximation ")
    print(f"of New_M by an invertible Mueller matrix. This is the final matrix obtained by the ECM.")
    print("------------------------------------------------------------------------------------")
    m_appr = appr_invert_mueller_matrix("New_M", Matrix(np.around(np.array(new_m).astype(np.float64), decimals=8)))
    return m_appr


new_m_0 = request_matrix("M")
print('The matrix M is: \n', np.array(new_m_0).astype(np.float64))
ecm(new_m_0)

\end{lstlisting}

\bigskip

\noindent \textbf{Acknowledgment. }We wish to express our gratitude to
Felipe Meneses for his help in obtaining valuable bibliographical material.

\noindent \textbf{Funding}: Our study was partially funded by a grant
AG200823 from the Direcci\'{o}n General de Asuntos del Personal Acad\'{e}%
mico, of Universidad Nacional Aut\'{o}noma de M\'{e}xico.

\end{document}